# Numerical investigation of stability of low-current needle-to-plane negative corona discharges in air

N. G. C. Ferreira[1,2], P. G. C. Almeida[1,2], A. Eivazpour Taher[1,2,†], G. V. Naidis[3], and M. S. Benilov[1,2,*]

[1]Departamento de Física, FCEE, Universidade da Madeira, Largo do Município, 9000 Funchal, Portugal

[2] Instituto de Plasmas e Fusão Nuclear, Instituto Superior Técnico, Universidade de Lisboa, 1041 Lisboa, Portugal

[3]Joint Institute for High Temperatures, Russian Academy of Sciences, Moscow 125412, Russia

[†]Present address: PLASMANT, Department of Chemistry, University of Antwerp, Antwerp, Belgium

[*]Corresponding author, email: benilov@staff.uma.pt

## Abstract

Negative DC corona discharges are known for their self-pulsing regime: the Trichel pulses. In some works, pulsed regimes, stochastic or periodic, have been observed immediately upon the inception of the discharge, while in other works the discharge was found to be ignited in a stedy-state (pulseless) mode, with the Trichel pulses developing at higher voltages. Recent theoretical and modelling work showed that the stationary negative corona between concentric cylinders in atmospheric-pressure air is stable immediately after the ignition. The pulseless mode was found also in the modelling of the needle-to-plane geometry, however in a quite narrow voltage range. This work studies conditions for a pulseless negative corona discharge in a needle-to-plane geometry to occur over a wide range of voltages, which will facilitate its unambiguous observation in the experiment. After the negative corona loses stability, the current evolution shows, after a small region of quasi-harmonic oscillations, pulses. These can be of small amplitude or regular Trichel pulses, which develop via standing-wave or ionization-wave mechanisms. Modelling results agree with available experimental data, both for the current-voltage characteristics and the stability limit of the pulseless negative corona discharge. An insight is given into stochastic Trichel pulses, which have been observed in experiments under certain conditions.

# 1 Introduction

Experimental investigation of the ignition of DC corona discharges started many years ago, e.g. [1–3]; works [4–10] may be cited as recent examples.

As far as negative corona is concerned, it was found in some works that it is ignited in a stationary (pulseless) mode; e.g., [1, 5, 6, 11, 12]. In [1], the pulseless regime was observed in the picoampere current range. In more recent experiments on DC negative coronas [5, 6, 11, 12], a steady (pulseless) discharge, denominated Townsend discharge, was observed before the appearance of the first Trichel pulses at currents below one or few microamperes. The review [13] indicates that the Trichel pulses occur in the negative pin-to-plane corona at currents above 1 $\mu$A, and a Townsend discharge occurs at lower currents. With increasing current (voltage), the pulseless regime gives way to stochastic pulses which, at still higher voltages, give way to periodic regime with Trichel pulses, characteristic of negative DC corona discharges; e.g., [1, 6].

On the other hand, some other works reported observations of pulsed regimes immediately upon the inception of DC coronas: stochastic pulses or stochastic bursts of a few pulses were observed in [3, 4], periodic Trichel pulses appear to have been observed in [9], and the intermittent regime was observed in [10]: at low voltages, transitions of the discharge from the Trichel pulse mode to the pulseless mode and back can occur.

There is a number of works in which the ignition of corona discharges is studied theoretically and/or numerically; works [14–20] may be mentioned as recent examples. The specific question of under what conditions, if any, DC corona discharges are ignited in pulsed or pulseless mode has begun to be studied in the last couple of years. A theory of stability of stationary low-current DC discharges was developed in [21]. The theory employed the fact that the ignition of self-sustaining DC discharges identifies in mathematical terms as a bifurcation of stationary solutions and general trends of the bifurcation theory were used to obtain hints on the stability of the bifurcating solutions. The theoretical conclusion has been tested numerically for a negative corona between concentric cylinders; a convenient test case for the theory. The atmospheric-pressure air was considered. Two independent approaches have been used: (1) study of linear stability against infinitesimal perturbations with the use of an eigenvalue solver and (2) following the time development of finite perturbations with the use of a time-dependent solver. The theory, the linear stability analysis, and the time-dependent modelling all show that the stationary negative corona is stable, i.e., the discharge is pulseless, immediately after the ignition. The range of voltages over which the discharge is stable is quite narrow in this geometry, just 8 V. Stability is lost at higher voltages: harmonic oscillations appear and their amplitude grows until a pulsed Trichel mode is attained.

These conclusions are consistent with results of the time-dependent modelling of low-current regimes of negative coronas in the needle-to-plane geometry [22, 23], except that the stability range found in [22, 23] was slightly wider, about 10 V. Similarly, the above conclusions are consistent with results of the time-dependent modelling of negative corona discharges in the needle-to-plane geometry reported in the work [24], although the stability range in [24] was not determined.

A number of interesting directions arise for future works, e.g., exploring the mechanism of the loss of stability and the potential effect of negative ions in this mechanism.

However, the first step is to find an unambiguous experimental confirmation of the theoretical prediction of negative corona discharge being steady-state (pulseless) immediately upon the ignition; the existing experimental information on this point is contradictory, as discussed above. Given that corona discharge voltages are several kilovolts or higher, detecting pulseless regimes that occur in a range of discharge voltages as narrow as 10 V is hardly realistic. Therefore, it is necessary to try to find conditions under which a pulseless negative corona discharge occurs over a sufficiently wide range of voltages upon the ignition, significantly wider than 10 V, in order to facilitate its unambiguous observation in experiment. This is one of the goals of this work. The needle-to-plane geometry is considered. Another goal is to provide insight into stochastic Trichel pulses, which have been observed in experiments under certain conditions as discussed above.

The paper is organized as follows. The numerical model of low-current corona discharges in atmospheric-pressure air, which is used in this study, and the method for studying stability are briefly described in section 2. In section 3, modelling results on stability of the negative corona immediately after the ignition against small perturbations are reported and discussed, and the temporal evolution of perturbations is analyzed. Stability against finite perturbations and appearance of, apparently, stochastic Trichel pulses are studied in section 4. Section 5 is concerned with the comparison between modelling results and available experimental data. Conclusions are summarized in section 6.

## 2  Model, simulation conditions, and numerics

Modelling results reported in this work refer to atmospheric-pressure dry air. The numerical model of low-current discharges in high-pressure dry air, described in [25], was employed. The model comprises equations of conservation and transport of charged particles and the Poisson equation. The charged particles taken into account are the electrons, an effective species of positive ions, and three species of negative ions: $O^-$, $O_2^-$, and $O_3^-$. The transport equations for the charged particles are written under the drift-diffusion approximation. Transport and kinetic coefficients are treated as known functions of the local electric field. Also included in the model are differential equations describing the photoionization rates in the framework of the three-exponential model [26]. The discharge-induced heating and the convective motion of the neutral gas are negligible at small discharge currents considered in this work, and therefore the gas temperature is set equal to 300 K and the convective transport of the ions and the electrons is neglected.

The simulations were performed in a needle-to-plane geometry with axial symmetry. The computation domain is schematically shown in figure 1. Nine discharge configurations indicated in table 1 have been considered. Here $d = |OA|$ is the discharge gap. In the configurations 1-8, the needle had the shape of a cylindrical rod ending with a conic section with a spherical tip of radius $R$. The cylindrical section had the radius $|BC| = 255\ \mu m$ and the length of 35 mm, the length of the conic section including the tip was 3 mm. The radius of the anode was $|KO| = 40$ mm in configurations 1, 3, 5, 7 and $|KO| = 80$ mm in configurations 2, 4, 6, 8. In the configuration 9, the needle had the shape of a cylindrical rod ending with a hemispherical tip. The cylindrical section had the radius $|BC| = 125\ \mu m$ and the length of the needle was 30 mm including the tip. The radius of the anode was

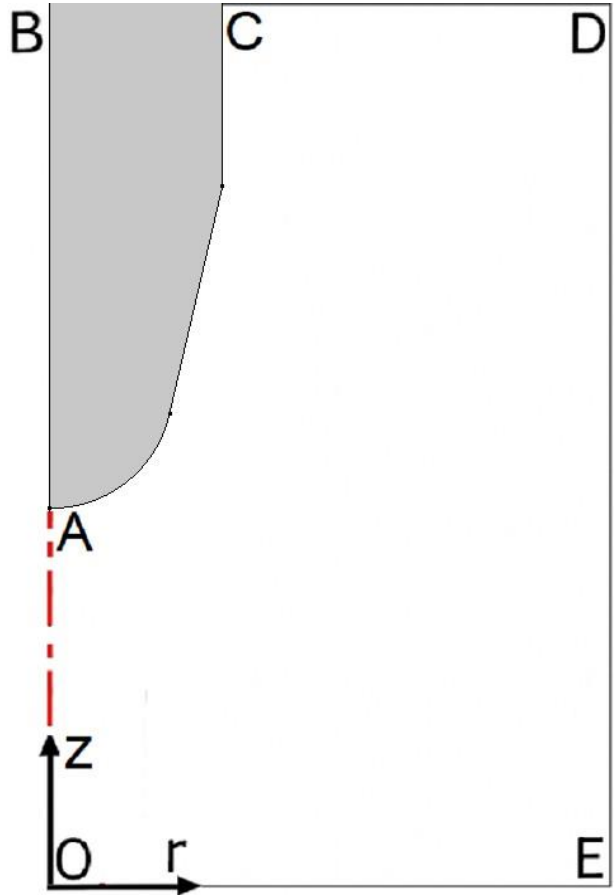

Figure 1. Schematic of the computation domain. Shaded area bounded by the contour ABC is the needle cathode. EO is the surface of the disc anode. CDE is the computational boundary.

|KO| = 100 mm.

Standard boundary conditions were applied at the surfaces of the electrodes, AC and KO; see, e.g., the discussion in [25]. The coefficient of secondary electron emission from the needle was $2 \times 10^{-4}$. The boundary conditions at the computational boundary CDK were zero normal derivatives of the charge-particle densities and electrostatic potential.

The boundary-value problem was solved with the use of stationary or time-dependent solvers of commercial software COMSOL Multiphysics®. The numerical model is implemented in a way to ensure the conservation of electric current, therefore, there was no need for a specific method for evaluating the discharge current, as done for example in [27–30] using the Sato equation [31].

For each discharge geometry, the computations have been performed in two steps. First, the inception voltage and all possible stationary states of the discharge, regardless of their stability, are computed by combining the resonance method and a stationary solver as described in section IV A of [25]. After that, stability of different states is investigated by applying a perturbation to the corresponding stationary solution and following the development of the perturbation by means of a time-dependent solver. If the perturbation decays in the course of temporal evolution and the discharge relaxes to the original (unperturbed) stationary state, then this state is stable against the considered perturbations. Otherwise, the state is unstable.

| Config-uration | $R$ ($\mu$m) | $d$ (mm) | $U_0$ (V) | $U_C - U_0$ (V) | $I_C$ (nA) |
|---|---|---|---|---|---|
| 1 | 2 | 10 | 1735 | 1460 | 2803 |
| 2 | 2 | 50 | 1988 | 5977 | 3882 |
| 3 | 10 | 10 | 1746 | 179 | 167 |
| 4 | 10 | 50 | 2002 | 857 | 170 |
| 5 | 20 | 10 | 1867 | 56 | 52 |
| 6 | 20 | 50 | 2144 | 320 | 51 |
| 7 | 100 | 10 | 3004 | 15 | 28 |
| 8 | 100 | 50 | 3365 | 119 | 26 |
| 9 | 125 | 20 | 3459 | 37 | 29 |

Table 1. Range of stability of negative corona for different discharge configurations against perturbations with $\Delta U = -1$ V. $R$: radius of the tip of the needle. $d = |OA|$: discharge gap length. $U_0$: inception voltage. $U_C$, $I_C$: discharge voltage and current at the last stable state. $U_C - U_0$: range of stable discharge.

Let $U$ be the applied voltage characterizing the stationary state of which the stability we want to investigate. The stationary solution for the voltage equal to $U + \Delta U$ with a value of $\Delta U$ equal to one, a fraction of one, or few volts was used as an initial condition for the time-dependent solver. This means the perturbation was set equal to the difference between the stationary solutions for the voltages $U + \Delta U$ and $U$.

## 3 Stability against small perturbations

This section is concerned with the study of stability of negative corona discharge against small perturbations, with $\Delta U = -1$ V unless otherwise specified.

### 3.1 Currents-voltage characteristics and limit of stability of stationary negative corona

Typical modelling results are shown in the figure 2. The inception voltage in this example is $U_0 = 1867$ V. Upon the ignition, stationary states of the negative corona discharge are stable up to the voltage $U_C = 1923$ V (corresponding to $I_C = 52$ nA; state $C$ in the figure 2). The stationary states are unstable and a pulsed regime occurs for higher discharge voltages, $U > U_C$. If the voltage keeps getting increased, the stability is eventually regained, which happens at $U = 8230$ V and $I = 47$ $\mu$A (state $D$ in the figure 2), and the steady-state discharge occurs again.

The stationary negative corona discharge being stable immediately after the ignition was predicted by the theory [21]. The loss of stability occurring at the state $C$ was predicted by the linear stability analysis [21] and is consistent with results of the modelling [21–24]. The recovery of stability occurring at the state $D$ is well known from both the experiment and the modelling; e.g., [4–6, 8, 11].

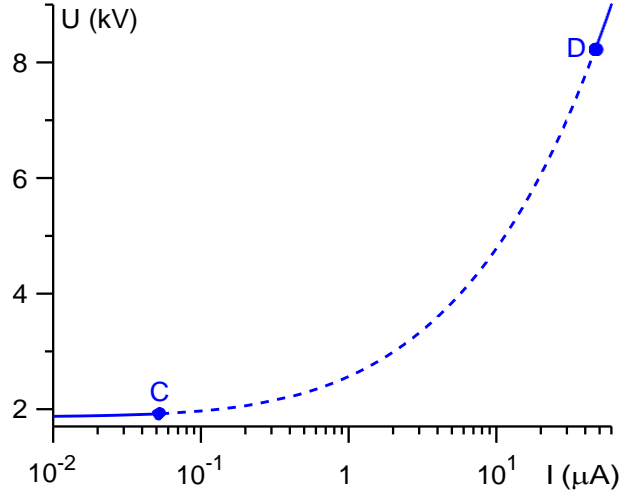

Figure 2. Current voltage characteristic (CVC) of the negative corona discharge given by the steady-state modelling. Discharge configuration 5. Solid lines: stable states. Dashed line: unstable states. Points $C$ and $D$: states where the stability changes.

The effect of variation of the tip radius $R$ and the discharge gap length on the discharge voltages and on the range of stable discharge is illustrated by table 1 and figure 3. For higher values of $R$ the discharge voltages are higher, as expected. The stability range increases as $R$ is decreased. The computed current-voltage characteristics (CVCs) for needles with $R = 2\,\mu$m and $10\,\mu$m are almost coincident (blue and red lines, respectively, in figure 3) and the CVC for the needle with $R = 20\,\mu$m is also rather close (the green line). In other words, the exact value of the tip radius of sharp needles of this geometry does not affect strongly the CVCs of the stationary corona, which is understandable since the conic angle is virtually the same for small $R$ values. On the other hand, the effect of the tip radius on the stability is much stronger, as evidenced by the data in the last two columns in table 1 and by the blue, red, and green circles in figure 3.

In order to give an idea of the effect of variation of the conic angle at a fixed tip radius, simulations with a length of the conical section of the needle of 2 mm, instead of 3 mm as above, were performed for configurations 2 and 4. The inception voltage increased by 136 V and 127 V, respectively, i.e., by just 6 or 7%. On the other hand, the range of stability was reduced by 1474 V and 188 V, respectively, i.e., by 25 and 28%. Again, a variation of the geometry of a sharp needle produces a stronger effect on the stability of the stationary corona than on its CVC.

An increase in the gap length causes an increase in the discharge resistance and, consequently, in the discharge voltage; a well-known effect. Except in the case $R = 2\,\mu$m, the current at the stability limit, $I_C$, is approximately the same for both discharge gaps 10 and 50 mm. The increase in the gap length causes an increase also in the stability range in terms of voltages, $U_C - U_0$, which is again a consequence of the increase in the discharge resistance.

The above results concern the stability of negative corona discharge against pertur-bations with $\Delta U = -1$ V. The effect of somewhat bigger perturbations on the stability range is exemplified in table 2. The increase of the amplitude of the perturbation $|\Delta U|$ from 1 to 3 V causes a decrease of the stability range as expected, however this decrease is small, about 10%. A further increase of the amplitude of the perturbation $|\Delta U|$ from 3 to 5 V does not visibly affect the stability range.

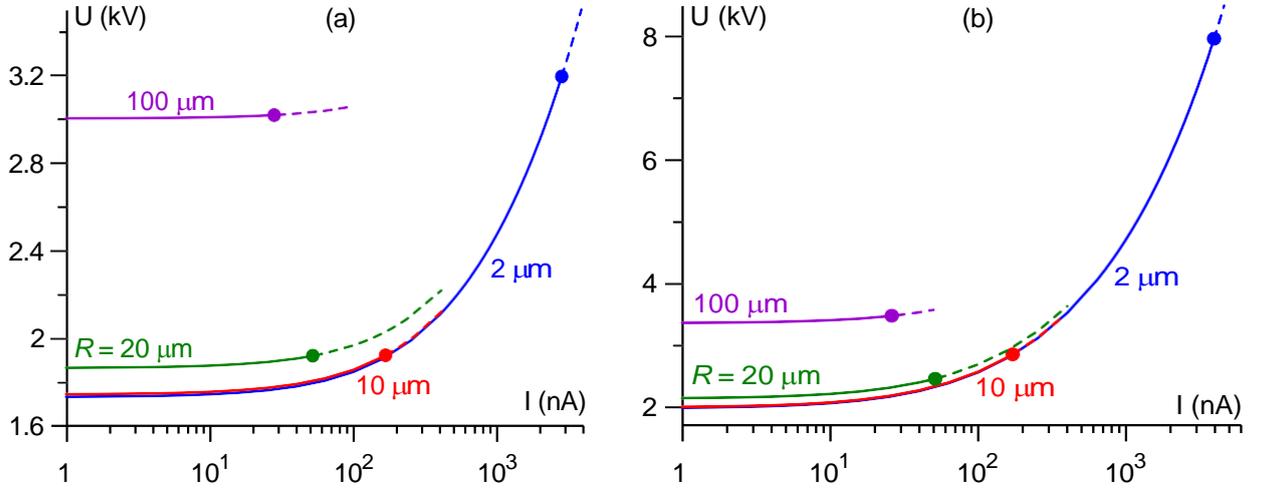

Figure 3. Effect of the tip radius on the stability limit of the negative corona. (a) Configurations 1,3,5 and 7, discharge gap of 10 mm. (b) Configurations 2,4,6 and 8, discharge gap of 50 mm. Solid lines: stable regions. Dashed lines: unstable regions.

| $\Delta U$ (V) | $U_C - U_0$ (V) | $I_C$ (nA) |
|---|---|---|
| −1 | 56 | 52 |
| −3 | 51 | 46 |
| −5 | 51 | 46 |

Table 2. Effect of the perturbation amplitude over the stability range of negative corona in configuration 5 ($R = 20\,\mu$m and $d = 10$ mm).

One can conclude that an increase of the nonuniformity of the electric field, whether it is produced by decreasing the needle tip radius and/or conic angle for the same gap or by increasing the gap length for the same needle shape, causes an increase in the range of stability of the quasi-stationary negative corona. The dependence of the stability range on the discharge geometry may be rather strong, especially for sharp needles. For the conditions considered, the range of stability varies significantly, from nearly 6 kV and 4 $\mu$A for configuration 2 to 15 V and 28 nA for configuration 7. Variations of the amplitude of perturbations, $|\Delta U|$, by a few volts do not significantly affect the range of stability, in agreement with the modelling [21].

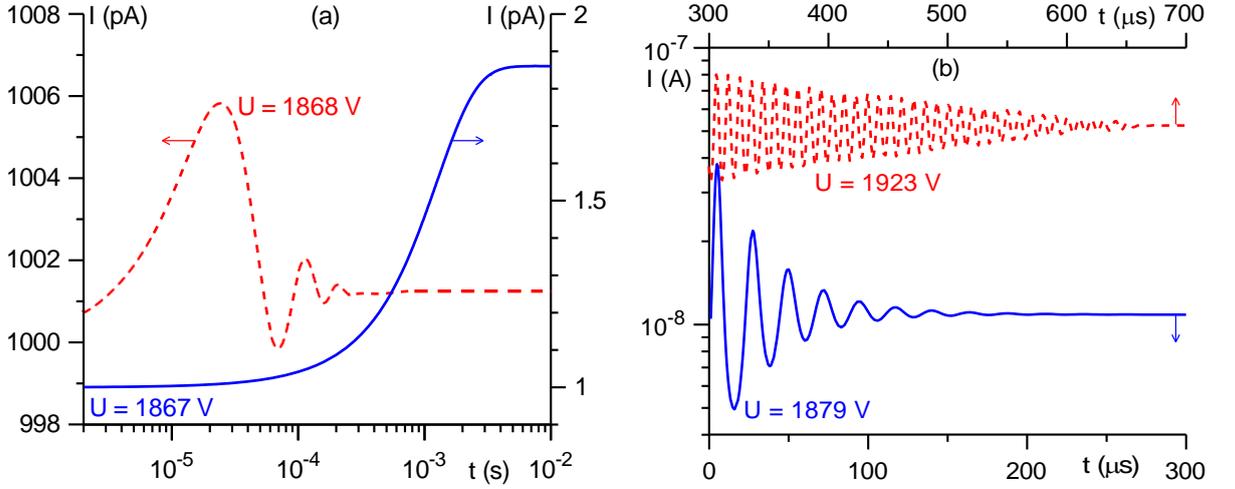

Figure 4. Computed evolution of current for perturbations of four stable states of the negative corona in configuration 5. (a) perturbation $\Delta U = -1$ mV. (b) $\Delta U = -1$ V.

## 3.2 Temporal evolution of perturbations and the outcome of the instability

The computed evolution of current for perturbations of four stable states of the negative corona in configuration 5 (the tip radius $R = 20\,\mu$m and the gap length $d = 10$ mm), $U = 1867$, 1868, 1879, and 1923 V is shown in figures 4(a) and (b), for imposed perturbations $\Delta U = -1$ mV and $-1$ V, respectively. The solid line in figure 4(a), corresponding to $U = 1867$ V, shows current growing monotonically with time until a steady state is reached in about 4 ms. For the state just 1 V above ($U = 1868$ V), shown by the dashed line, current grows in an oscillatory manner until a steady state is reached in about 0.5 ms. Given that the perturbation in both cases was the same, $\Delta U = -1$ mV, one can conclude that the character of evolution of discharge current, monotonic or oscillatory, is a characteristic of the stationary state itself and not of the imposed perturbation, provided that perturbations are small, as it should be according to the linear stability theory. Note that both monotonic and oscillatory current variations during the development of infinitesimal perturbations of stable states of a negative corona in a concentric-cylinder geometry were identified in [21] through linear stability analysis and numerical modeling similar to that used in this work, except that the modelling was one-dimensional (1D) in [21] and two-dimensional (2D) in this work.

The damped oscillations seen in figure 4(a) for the state $U = 1868$ V are seen also in figure 4(b) for both states $U = 1879$ and 1923 V. Note that the state $U = 1923$ V is the last stable state for configuration 5, i.e., state $U = 1924$ V is unstable. The period of the oscillations decreases with increasing voltage. The duration of the oscillation phase is of the order of a few hundred microseconds.

The computed evolution of current for perturbations of two unstable states is shown in figure 5. It is seen that for the state $U = 1924$ V, which is just above the limit of stability (i.e., this state corresponds to the voltage of 1 V above the last stable state), quasi-harmonic oscillations with the amplitude of approximately $0.1\,\mu$A and the frequency of around 100 kHz occur; figure 5(b). With time, the oscillations transition to narrow periodic pulses with the amplitude of around 0.4 $\mu$A and the frequency of around 70 kHz, as shown by figure 5(c) and by the solid line in figure 5(d). The transition is very slow, which is understandable since the growth increment of perturbation is small in the vicinity of the stability limit.

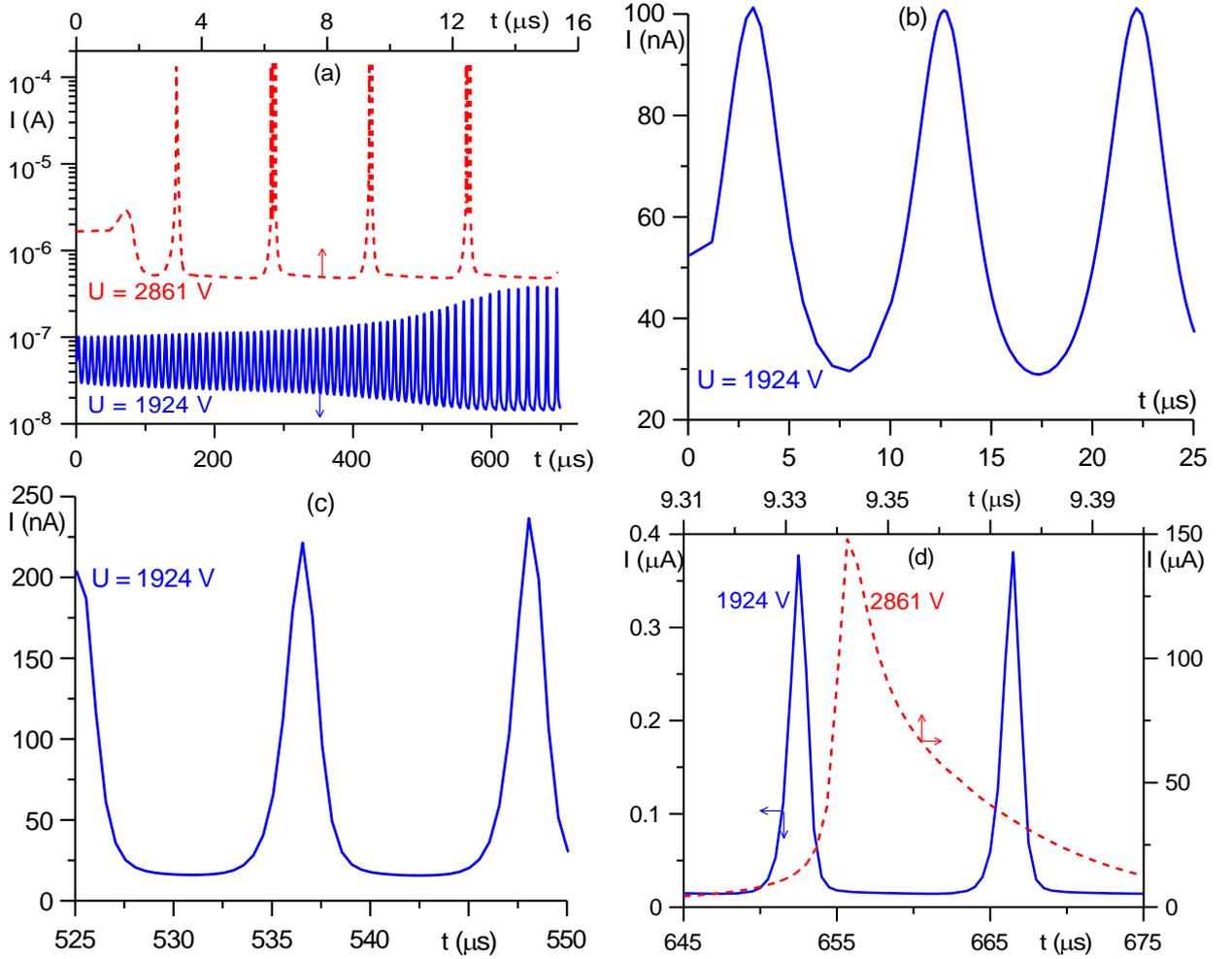

Figure 5. The computed evolution of current for perturbations of two unstable states of the negative corona in configuration 5. Solid: $U = 1924$ V. Dashed: $U = 2861$ V.

For the state $U = 2861$ V, there is just one low-current oscillation (the maximum current of about 3 $\mu$A) which gives way to periodic pulses with the amplitude of around 150 $\mu$A and the frequency of around 320 kHz. Details of the shape of the pulses are shown by the dashed line in figure 5(d). Note that the results shown in figure 6(a) of [21] revealed several low-current quasi-harmonic oscillations before the well-developed periodic pulses of high amplitude have appeared, instead of just one low-current oscillation as seen for $U = 2861$ V in figure 5(a). This is because the state to which figure 6(a) of [21] refers is very close to the stability limit. Note that the simulation results for conditions of figure 5(a) for a state at a significantly lower voltage, $U = 2051$ V, which are skipped for brevity, revealed several low-current quasi-harmonic oscillations, in qualitative agreement with the results shown in figure 6(a) [21].

A pulsed regime in negative corona discharges is well-known; the so-called Trichel pulses. The parameters of the Trichel pulses vary over a wide range depending on the discharge conditions (e.g. [32–37]): amplitudes from hundred of microamperes to a few milliamperes, frequencies from hundreds of kilohertz to a few megahertz, rise times from a few nanoseconds to, say, 20 ns, fall times longer by a factor of typically two to few tens. The periodic pulses that develop in the state $U = 2861$ V and are exemplified by the dashed line in figure 5(d) possess all hallmarks of the Trichel pulses. In contrast, the pulses that develop in the state $U = 1924$ V, which is close to the stability limit, and are shown in figure 5 by the solid lines, do not fit into the above ranges: amplitude of around $0.4\,\mu$A, frequency of around 70 kHz, and approximately equal rise and fall times of around 2.5 $\mu$s. Hence, these pulses are different from the Trichel pulses. For brevity, these pulses may be termed low-current ones.

The evolution of axial distributions of the discharge parameters during the develop- ment of the instability shown in figure 5 is illustrated by figure 6. $n_{A+}$ is the density of effective species of positive ions [25], and $n_{A-}$ is the sum of densities of the three species of negative ions ($O^-$, $O_2^-$, and $O_3^-$). Figures 6(a) and (b) refer to, respectively, the first maximum and the first minimum of current, shown in the figure 5(b). Figures 6(c) and (d) refer to, respectively, the maximum and minimum of a developed periodic pulse, shown by the solid line in the figure 5(d). Figures 6(e), (f), (g), and (h) refer to, respectively, the first maximum, the first minimum, the second maximum, and the second minimum of current, shown by the dashed line in the figure 5(a).

$E$ in figure 6 is the axial component of the electric field, which in this case equals the electric field modulus, hence $E/N$ has the meaning of the reduced electric field as before. The dotted lines in figure 6 represent the applied (non-distorted) reduced electric field. Note that the spatial increase or decrease of the field distortion is not directly related to the sign of the space-charge density, which is due to 2D effects. Note also that the reduced electric field at the needle exceeds 1500 Td in all the cases. Since the electron distribution function become strongly anisotropic at fields that high, this range is beyond the region of applicability of the drift-diffusion local-field model, hence the computation results can only be qualitatively correct here.

In both cases $U = 1924$ and 2861 V, the density of the positive ions is more or less constant up to approximately 20 $\mu$m from the needle (cathode). The densities outside this region are significantly lower.

Figures 6(g) and (h), referring to the developed Trichel pulses, fit the well-known pattern of Trichel pulses. In particular, there is a significant distortion of the applied electric field. The applied field is enhanced near the needle and a structure resembling a space-charge sheath bounding the quasi-neutral plasma exists at the current maximum as seen in figure 6(g). The applied field is weakened near the needle at the current minimum as seen in figure 6(h).

The distributions of the discharge parameters at the beginning of the formation of the Trichel pulses, shown in figures 6(e) and (f), are substantially different: the charged-particle densities are substantially lower and there is no quasi-neutrality at any part of the gap; the applied field is weakened near the needle not only at the current minimum as seen in figure 6(e), but at the current maximum as well as seen in figure 6(f).

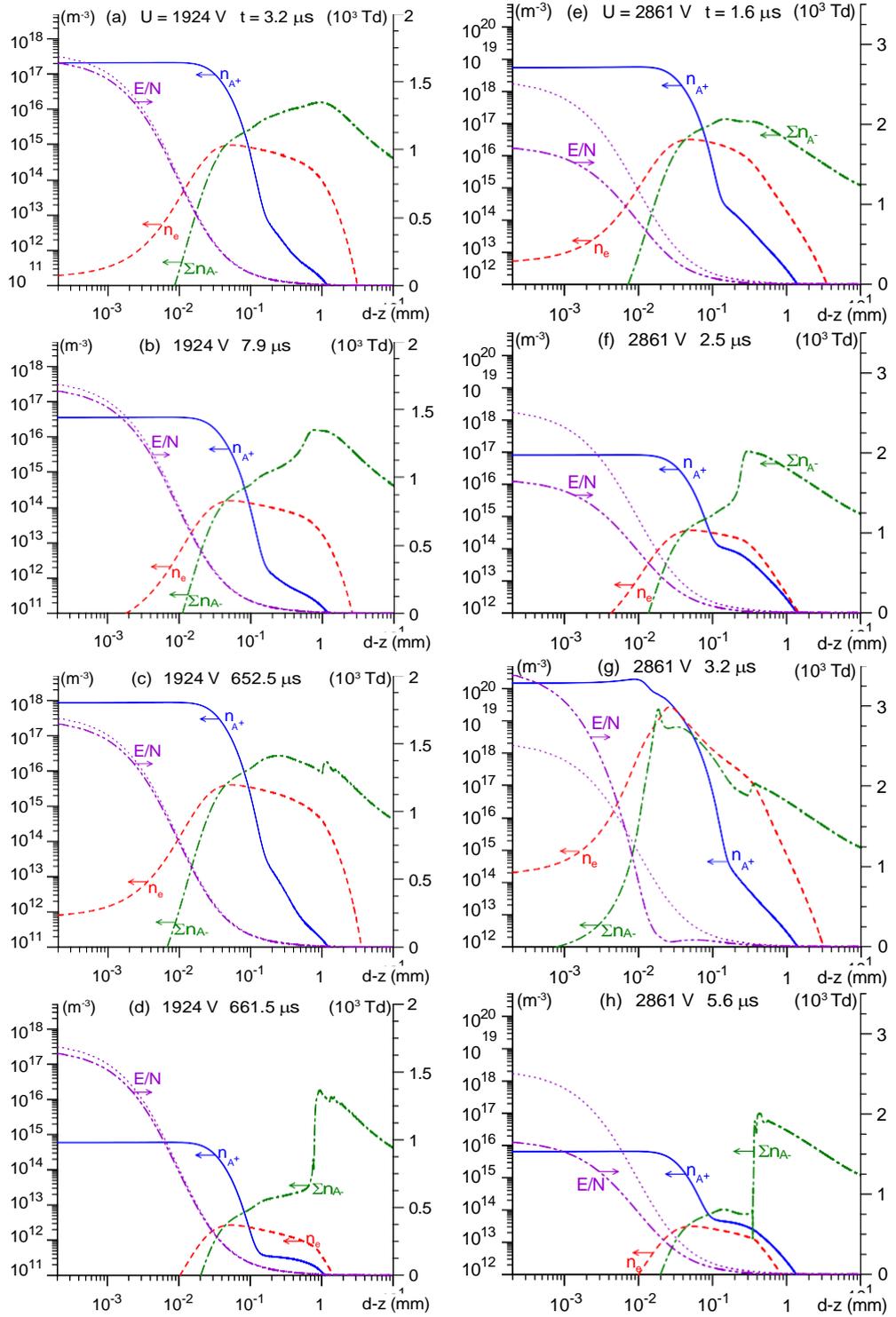

Figure 6. Evolution of axial distributions of discharge parameters during the development of the instability of two unstable states in configuration 5. The needle on the left, anode on the right. $n_{A+}$: density of an effective species of positive ions, $n_{A-}$: sum of densities of the three species of negative ions. Dotted: applied (non-distorted) reduced electric field. (a)-(d): $U = 1924$ V. (e)-(h): $U = 2861$ V.

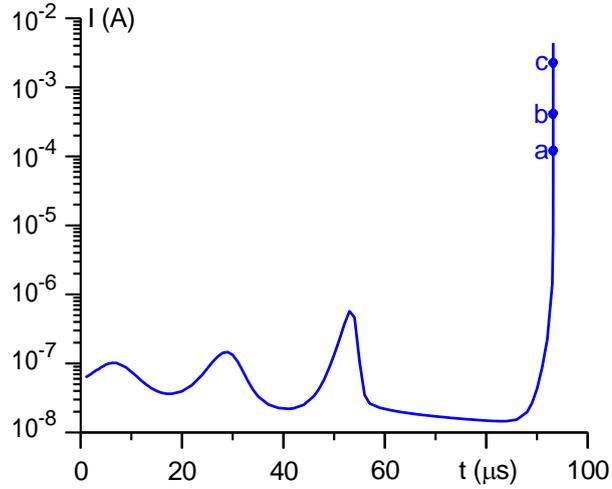

Figure 7. The computed evolution of current for perturbation of an unstable state under configuration 7. $U = 3040$ V. The circles mark instants to which figures 8(a) to (c) refer.

In the case of low-current pulses, shown in figures 6(a)-(d), the charged-particle densities are still much lower. The distortion of the electric field is weak. The applied field is somewhat weakened near the needle both at the current minima, as seen in figures 6(b) and (d), and the current maxima, as seen in figures 6(a) and (c). It follows that the low-current pulses occur on the background of a virtually unchanged electric field and hence represent a weakly nonlinear phenomena, their shape being a result of accumulation of weak nonlinear effects over many pulses.

As mentioned above, the thickness of the region near the needle where the density of the positive ions is approximately constant is about 20 $\mu$m and does not change much during the development of the instability in both cases $U = 1924$ and 2861 V. The maximum of the electron density occurs not far away from the edge of the region where the density of the positive ions is approximately constant, i.e., at about 20 $\mu$m, and does not change much as well, as seen in figures 6(a)-(d) for $U = 1924$ V and figures 6(e)-(h) for $U = 2861$ V. In this sense, one can say that both the low-current pulses and the Trichel pulses represent standing waves in these conditions.

The figures 5 and 6 refer to the configuration 5 of table 1, $R = 20$ $\mu$m and $d = 10$ mm. The low-current pulses, that develop via slow amplification of weakly nonlinear oscillations at states immediately above the stability limit, represent standing waves also for all the other discharge configurations detailed in table 1.

In contrast, Trichel pulses in configurations with bigger needle tip radii do not develop as standing waves. As an example, let us consider figures 7 and 8, referring to the configuration 7, $R = 100$ $\mu$m and $d = 10$ mm. The discharge voltage is $U = 3040$ V, about 20 V above the last stable state.

There are three small-current oscillation cycles in 7. Spatial distributions of the discharge parameters at this stage are qualitatively similar to those shown in figures 6(a)-(d) and are skipped for brevity. The positive-ion density at each instant during the oscillations is approximately constant within about 50 $\mu$m from the needle. The spatial position of the maximum of the electron density is also about about 50 $\mu$m and does not change

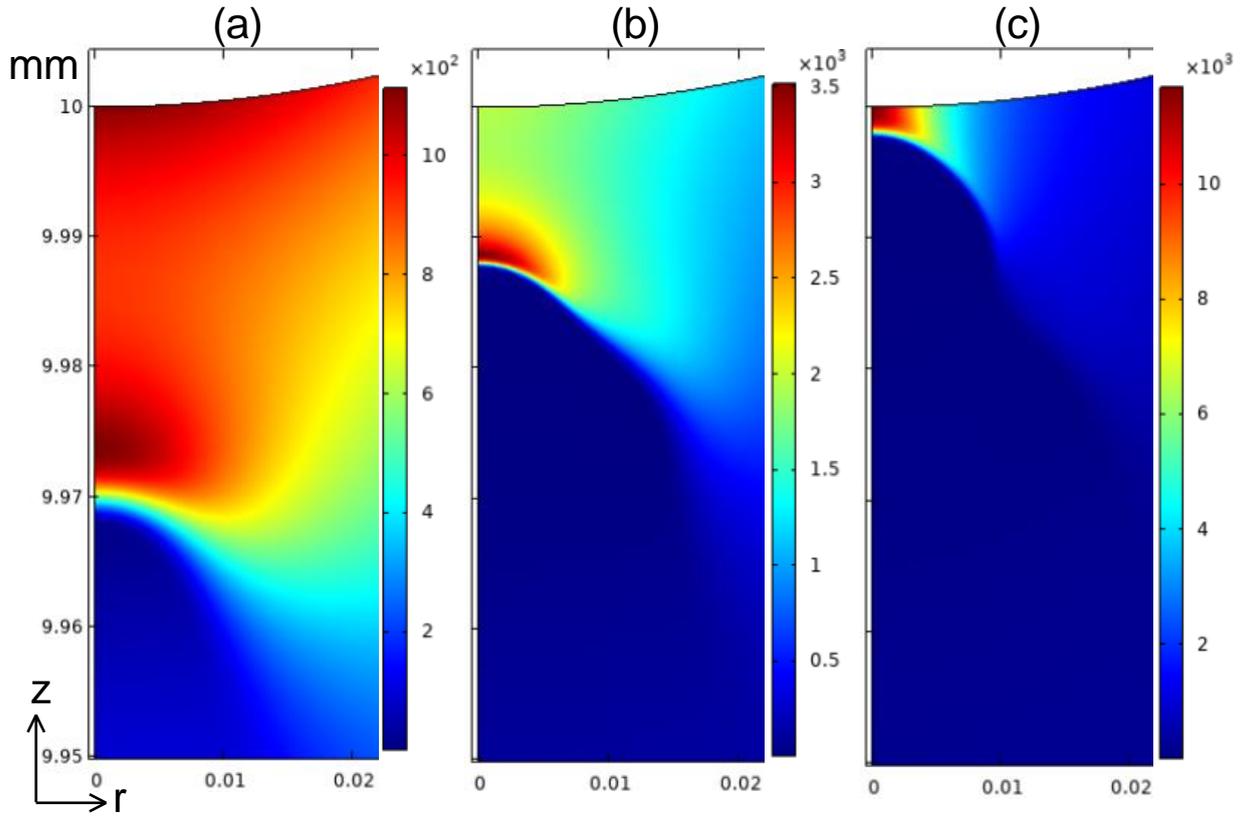

Figure 8. 2D distributions of the reduced electric field (in Td) in the vicinity of the needle for time instants shown in figure 7. (a): $t_a = 93.1712\,\mu s$. (b): $t_b = 93.1718\,\mu s$). (c): $t_c = 93.1719\,\mu s$. Spatial scales in mm.

much as well. One can view these oscillations as standing waves.

At about 93 $\mu$s, a maximum in the positive-ion density, accompanied by a distortion in the electric field, is formed at about 50 $\mu$m from the needle and starts shifting in the direction to the needle. Simultaneously, a steep increase of current begins: $I = 0.2\,\mu A$ for $t = 92\,\mu s$, $I = 1.4\,\mu A$ for $t = 93\,\mu s$, and $I = 7.7\,\mu A$ for $t = 93.15\,\mu s$. The absolute values of the charged particle densities rapidly increase, a space-charge sheath bounding quasi-neutral plasma is formed, and a maximum of the electric field appears: a cathode-directed ionization wave has been formed as seen in figure 8. Note that some authors use the term 'streamer' for such waves; e.g., [33]. Other authors, however, reserve this term for waves propagating in the direction of reduction of the intensity of the applied electric field, so this term is not used in this work.

As the ionization wave approaches the cathode, a very strong electric field, exceeding $10^4$ Td, is induced at a micrometer scale from the tip of the needle as seen in figure 8(c). This scenario is known in the literature for being extremely costly computationally, e.g. [38]; see also discussion in [39]. The drift-diffusion local-field model fails at fields that high and, additionally, field electron emission from the needle with enhancement of the electric field on surface non-uniformities may come into play. On the other hand, detailed information on the advanced nonlinear stage of instability development, although interesting in itself, adds little to the understanding of the stability of stationary states of negative corona discharges, on which this work is focussed. For these reasons, results reported in this work are limited to this stage. Let us only note that trial simulations have shown that the instability development in this case results in the formation of Trichel pulses with a mechanism involving ionization waves; an intuitively clear result.

The above results refer to configuration 7 of table 1, $R = 100\,\mu$m and $d = 10$ mm. The ionization wave develops also in configuration 8, where the tip radius is the same, $R = 100\,\mu$m, but the gap length is much bigger, $d = 50$ mm.

One can conclude that the development of the instability of stationary states of negative corona discharge for $U > U_c$ can lead to three possible outcomes: low-current pulses, Trichel pulses with a standing-wave mechanism, and Trichel pulses with a mechanism involving ionization waves. Note that these conclusions agree with the results of numerical modelling reported in section IV of [21] and, in particular, with the conclusion that the Trichel pulses from the theoretical point of view represent a result of development of harmonic oscillations, through the growth of which the stability of the negative corona discharge for $U > U_c$ is lost.

The low-current pulses occur immediately after the stability limit, i.e., for voltages exceeding $U_c$. At still higher voltages, the low-current pulses transition to the Trichel pulses. The voltage range where the low-current pulses occur may be very narrow; e.g., 2 and 25 V for configurations 7 and 8, respectively.

Trichel pulses with a standing-wave mechanism and those with the ionization waves have been obtained in previous simulations; e.g., [9, 24, 34, 36, 37, 40–43] and [29, 38, 44–46], respectively. The formation of ionization waves has been observed in the experiments [47, 48]. The low-current pulses apparently have not been obtained in previous simulation works. These pulses virtually do not disturb the applied electric field and their shape is a result of accumulation of weak nonlinear effects over many cycles. Note that such pulses appear to have been observed in recent experiments [10]; see discussion in section 6. It would be very interesting to study these pulses in more detail.

As a side note, it is interesting to point out the following. According to the modelling of this work, standing-wave Trichel pulses are favoured by smaller needle tip radii and the opposite is true for Trichel pulses with ionization waves. This is consistent with results of most previous simulations: the Trichel pulses with standing waves were obtained in [9, 24, 34, 36, 37, 40, 42] for the needle radii in the range of 2 — 60 $\mu$m, while the Trichel pulses with ionization waves were obtained in [29, 45, 46] for the needle radii in the range of 80 — 400 $\mu$m (and in [44] for a sphere-to-plane negative corona with the sphere radius of 5 mm). However, there are also works [41, 43], where the Trichel pulses with standing waves have been obtained in the modelling for the needle radii of 250 and 270 $\mu$m, respectively. Presumably, this contradiction may stem from differences in the shape of the needle tip.

## 4 Finite perturbations: stochastic Trichel pulses

As shown by the modelling of section 3.2 above and of section IV [21], the periodic Trichel pulses from the theoretical point of view represent a result of development of small harmonic oscillations, through the growth of which the stability of the negative corona discharge for $U > U_c$ is lost. On the other hand, it is known from the experimental works cited in the Introduction that periodic Trichel pulses are in many cases preceded by stochastic (irregular) Trichel pulses.

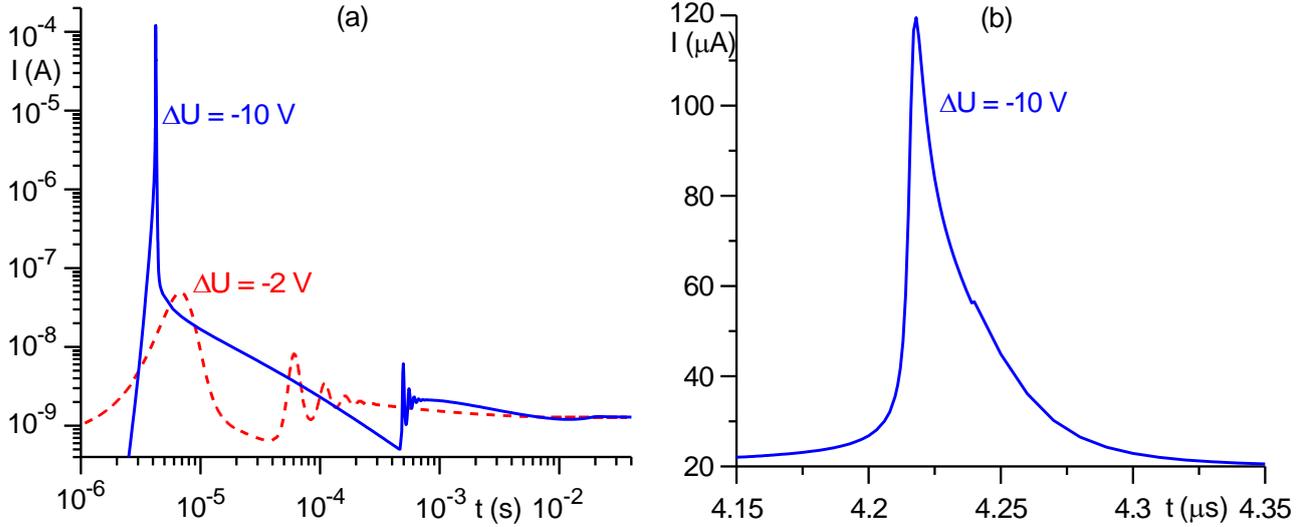

Figure 9. Computed evolution of two finite perturbations of a stable state of the negative corona discharge. Configuration 6, $R = 20\,\mu$m, $d = 50$ mm, state $U = 2154$ V. (a) Evolution on a wide time range, until the perturbations have decayed and the unperturbed stationary state has been recovered. (b) Magnification of figure (a) in the time range where the Trichel pulse occurs.

It can be hypothesized that the stochastic Trichel pulses represent a result of development of random finite perturbations of stable states. The computed evolution of current for two different finite perturbations of a stable state of the negative corona discharge is shown in figure 9 for configuration 6 ($R = 20\,\mu$m, $d = 50$ mm) and the stationary state $U = 2154$ V. The perturbation caused by the voltage variation $\Delta U = -2$ V, which is shown in 9(a) by the dashed line, decays in an oscillatory way until the unperturbed stationary state has been recovered. The maximum value of current, which is about 50 nA, is much smaller than values characteristic of the Trichel pulses. The perturbation caused by a much bigger voltage variation, $\Delta U = -10$ V, which is shown in figure 9(a) by the solid line, first evolves into a typical Trichel pulse with the maximum current of around 120 $\mu$A, as seen in the magnification in figure 9(b). After the pulse, the current decays by about five orders of magnitude. Then a few low-current oscillations occur, which subsequently decay and the unperturbed state has been recovered.

This example shows that random variations of conditions of pulseless negative corona can provoke the appearance of solitary Trichel pulses, which appear to be stochastic, provided that these variations are not too small (voltage perturbations of 10 V are sufficient in this example). Thus, one can hypothesize that stochastic Trichel pulses are caused by finite (non-small) fluctuations of conditions of pulseless negative corona discharge.

Simulations of finite perturbations of unstable states close to the stability limit have also shown a solitary Trichel pulse, which appears to be stochastic. However, in this case the Trichel pulse is followed by low-current periodic pulses. The time of appearance and amplitude of these periodic pulses depend on the perturbation magnitude.

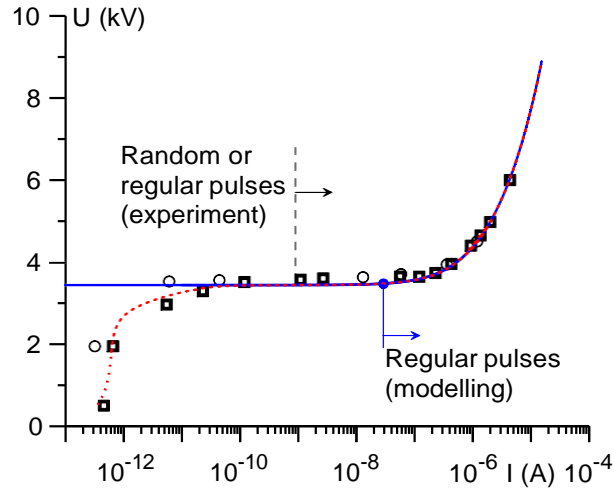

Figure 10. CVC and stability limit for negative rod-to-plane geometry in atmospheric pressure dried air. Rod tip radius of 125 $\mu$m and 20 mm discharge gap. Squares: experimental data obtained in [1] for increasing voltage. Circles: experimental data obtained in [1] for decreasing voltage. Solid: modelling. Dotted: see text.

## 5 Comparison with the experiment

There is a number of experiments [1, 5, 6, 11, 12] where a steady (pulseless) negative corona discharge was observed immediately after the discharge ignition, before the appearance of the first Trichel pulses. The geometry of the needle tip was indicated in only one of these works, the classic paper [1]. As shown in section 3.1, a variation of geometry of the needle tip can produce a strong effect on the stability of the pulseless discharge, hence the natural object for comparison are experimental data reported in [1].

Modelling results for conditions [1] are shown in the last line of table 1 and figure 10. Squares and circles in the figure represent points of the CVC obtained in the experiments [1] with increasing and decreasing voltages, respectively. The solid line represents the computed CVC of the stationary discharge. There is a good agreement between the computed and measured data in the current range $10\,\text{pA} < I < 10\,\mu\text{A}$.

The computed and measured discharge voltages are virtually constant in the current range $10\,\text{pA} < I < 100\,\text{nA}$. The discharge is self-sustained, but the density of the charged particles in the discharge gap is very low for currents that small so that the Laplacian electric field induced by the electrodes remains unperturbed. One can call this regime the Townsend discharge. For higher currents, the perturbation of the Laplacian electric field becomes non-negligible and the CVC $U(I)$ starts increasing, according to both the experiment and the modelling. If the discharge current is reduced from 10 pA to values below 1 pA, the measured discharge voltage significantly decreases, while the computed voltage remains constant. One should assume that there is an external ionization source in this experiment that can produce the maximum current of the order of 10 pA. In other words, the discharge is non-self-sustained for current of the order of 10 pA or lower, and it is unsurprising that the numerical model being used, which describes self-sustained discharges, becomes inapplicable for currents that low.

According to [49], an effective rate of the electron–ion pair appearance in air due to a normal concentration of radon is up to $10^7$ m$^{-3}$ s$^{-1}$ and the source rate of ionization caused by cosmic rays is in the order of $10^3$ m$^{-3}$ s$^{-1}$, i.e., much lower. Simulations with a constant external source term of $10^7$ m$^{-3}$ s$^{-1}$ in the equations of conservation of the positive ions and the electrons for $U = 3359$ V, which is by 100 V lower than the corona inception voltage $U_0$ indicated in the last line of table 1, gave the current of 7 fA, which is by three orders of magnitude lower than the above-mentioned current of 10 pA. (Note that there were no external source terms in the equations of conservation of negative ions, however the negative ions were produced through the attachment.) Hence, neither natural radioactivity nor cosmic radiation contributes appreciably to the external ionization effect seen in figure 10.

On the other hand, the negative corona discharge in the experiment [1] was triggered by ultraviolet illumination of the needle to liberate photoelectrons. No quantitative data on the photoemission current was given. For illustrative purposes, a modelling was performed with the photoemission current of 3 pA uniformly distributed along the needle surface; the dotted line in figure 10. Unsurprisingly, the agreement with the experiment is good.

The onset of Trichel pulses in the experiment occurs at $I \approx 1$ nA and the voltage which to the graphical accuracy coincides with the inception voltage. This onset current is marked in figure 10 by the dashed vertical line. As the voltage was raised, the Trichel pulses occurred more frequently and quickly became regular in frequency and amplitude.

In the modelling, the last stable state, i.e., the one before the appearance of regular low-current pulses, corresponds to the voltage $U_c = 3496$ V, i.e., exceeds the inception voltage by just 1%. The corresponding discharge current, $I_c \approx 29$ nA, is marked by the solid vertical line in figure 10. (Note that the values computed with the above-described account of photoemission are quite close: $U_c = 3500$ V, $I_c \approx 32.6$ nA.) Low-current pulses develop in a very narrow range of voltages (about 10 V) above $U_c$ and the Trichel pulses with ionization waves appear for higher voltages.

Thus, both in the experiment and the modelling the stability of pulseless negative corona is lost and the Trichel pulses develop at states with voltages very close to the inception voltage. Since the experimental and computed values of the inception voltage are very close to each other, the voltages at which the stability of pulseless negative corona is lost and the Trichel pulses develop in the experiment and the modelling are very close as well. There is a significant difference in the current values, which is unsurprising since the experimental value refers to the onset of the (random) Trichel pulses, while the value obtained in the modelling refers to the appearance of the low-current regular pulses. Moreover, small variations of the needle tip shape affect the stability range much stronger than the stationary CVC.

## 6 Summary and conclusions

The stability of low-current negative corona discharges in the needle-to-plane geometry in atmospheric-pressure air was investigated numerically under a wide range of discharge configurations, with the aim to find conditions under which a pulseless negative corona discharge occurs over a sufficiently wide range of voltages upon the discharge ignition, in order to facilitate its unambiguous observation in experiment. For each discharge geometry, the computations have been performed in two steps. First, the inception voltage and all possible stationary states of the discharge, regardless of their stability, are computed by means of a stationary solver. After that, stability of different states is investigated by applying a perturbation to the corresponding stationary solution and following the development of the perturbation using a time-dependent solver.

It is shown that, in agreement with the previous theory, linear stability analysis, and numerical modelling, the negative corona discharge is stable immediately after the ignition, i.e., can be operated in a pulseless mode. The discharge loses stability on the rising section of the CVC. The stability range increases as the value of the tip radius $R$ is decreased. The exact value of $R$ of sharp needles does not affect strongly the CVCs of stationary corona. On the other hand, the effect of $R$ on the stability is much stronger.

An increase of the nonuniformity of the electric field, whether it is produced by decreasing the needle tip radius and/or conic angle for the same gap or by increasing the gap length $d$ for the same needle shape, causes an increase in the range of stability of the quasi-stationary negative corona. For the conditions considered, the range of stability varies significantly, from nearly 6 kV and 4 $\mu$A for $R = 2\,\mu$m and $d = 50$ mm to 15 V and 28 nA for $R = 100\,\mu$m and $d = 10$ mm.

Perturbations of stable states immediately after ignition decay in a monotonic way. For states with slightly higher voltages, still within the stable region, perturbations decay in an oscillatory way. The character of evolution of discharge current is a characteristic of the stationary state itself and not of the imposed perturbation, provided that perturbations are small, as it should be according to the linear stability theory.

The time evolution of the perturbations of (unstable) stationary states with voltages $U$ slightly above the stability limit $U_c$ begins with the appearance of quasi-harmonic oscillations of current. These oscillations occur on the background of a virtually unchanged electric field and hence represent a weakly nonlinear phenomena. Being weakly nonlinear, the oscillations gradually develop low-current periodic pulses with amplitudes in the order from tenths to one microampere. The shape of the pulses is a result of accumulation of weak nonlinear effects over many cycles. The spatial position of the maximum of the electron density remains virtually unchanged during the oscillations, hence these oscillations may be viewed as standing waves.

As the difference $U - U_c$ increases (i.e., at stationary states further away from the stability limit), the time evolution of the perturbations begins with a few oscillations which rapidly evolve into the classic Trichel pulses, with the amplitudes in the order from hundred of microamperes to a few milliamperes.

For smaller values of $R$, the Trichel pulses represent standing waves, similarly to the low-amplitude current pulses. In contrast, the Trichel pulses in configurations with bigger values of $R$ develop not as standing waves but rather as moving ionization waves. This is consistent with the results of most previous simulations of the well-developed Trichel pulses. The low-current pulses apparently have not been obtained in previous simulation works; on the other hand, they resemble the ones observed in recent experiments [10]: in some cases, the negative corona was ignited in the form of periodic pulses with the amplitude of about $1 - 2\,\mu$A, alternating with a pulseless mode.

In summary, the development of the instability of stationary states of negative corona discharge for $U > U_c$ can lead to three possible outcomes: low-current pulses, Trichel pulses with a standing-wave mechanism, and Trichel pulses with a mechanism involving ionization waves. One can say that from the theoretical point of view, the Trichel pulses represent a result of development of harmonic oscillations, through the growth of which the stability of the negative corona discharge is lost for $U > U_c$.

It is shown that random variations of conditions of pulseless negative corona can provoke the appearance of solitary Trichel pulses, which appear to be stochastic, provided that these variations are not too small (voltage perturbations of 10 V are sufficient in the example shown in figure 9). Simulations of finite perturbations of unstable states with voltages slightly above the stability limit have also shown a solitary Trichel pulse, which appears to be stochastic and is followed, in this case, by low-current periodic pulses.

Modelling results for conditions [1] have shown a good agreement between the computed CVC of the stationary discharge and the measured data [1] in the current range $10\,\text{pA} < I < 10\,\mu$A. The stability of pulseless negative corona is lost and the Trichel pulses develop at states with voltages very close to the inception voltage both in the modelling and the experiment, and the values of these voltages are close to each other within the accuracy of the graphical data.

One can hope that the results of numerical investigation of stability of low-current negative corona discharges in the needle-to-plane geometry under a wide range of discharge configurations, reported in this work, will help to perform reproducible experimental observations of the pulseless low-current negative corona discharge, thus resolving contradictions in the available literature. It would be very interesting to investigate experimentally also the low-current pulses near the stability limit, that have been obtained in the modelling prior to the development of the regular Trichel pulses.

Acknowledgments The work at UMa/IPFN was supported by FCT—Fundação para a Ciência e a Tecnologia of Portugal under projects UIDB/50010/2020, UIDP/50010/2020, and LA/P/0061/2020. The work at the JIHT RAS was supported by the Ministry of Science and Higher Education of the Russian Federation (State Assignment No. 075-00269-25-00).